\documentclass[aps,prl,twocolumn,superscriptaddress,showpacs,preprintnumbers,amssymb]{revtex4-1}
\usepackage{graphicx }
\usepackage{dcolumn}
\usepackage{bm}
\usepackage{color}
\usepackage{amsmath}
\usepackage{amssymb}
\usepackage{ulem}

\newcommand{\nccf}{NaCaCo$_2$F$_{7}$}
\newcommand{\Tcw}{\Theta_{\rm CW}}
\newcommand{\mub}{\mu_{\mathrm{B}}}
\newcommand{\Tf}{T_{\mathrm{f}}}

\newcommand{\ertio}{Er$_{2}$Ti$_{2}$O$_{7}$}

\newcommand{\Tc}{T_{\mathrm{c}}}

\begin{document}

\title{
Spin freezing in the disordered pyrochlore magnet NaCaCo$_2$F$_7$:\\
NMR studies and Monte-Carlo simulations
}

\author{R. Sarkar}
\altaffiliation[]{rajibsarkarsinp@gmail.com}
\affiliation{Institut f\"ur Festk\"orperphysik, Technische Universit\"at Dresden, 01062 Dresden, Germany}
\author{J. W. Krizan}
\affiliation{Department of Chemistry, Princeton University, Princeton, NJ 08544, USA}
\author{F. Br\"uckner}
\affiliation{Institut f\"ur Festk\"orperphysik, Technische Universit\"at Dresden, 01062 Dresden, Germany}
\author{E. C. Andrade}
\affiliation{Instituto de F\'{i}sica Te\'{o}rica, Universidade Estadual Paulista, Rua
Dr. Bento Teobaldo Ferraz, 271 - Bl. II, 01140-070, S\~{a}o Paulo, SP,
Brazil}
\affiliation{Instituto de F\'{i}sica de S\~ao Carlos, Universidade de S\~ao Paulo, C.P. 369, S\~ao Carlos, SP,  13560-970, Brazil}
\author{S. Rachel}
\author{M. Vojta}
\affiliation{Institut f\"ur Theoretische Physik, Technische Universit\"at Dresden,
01062 Dresden, Germany}
\author{R. J. Cava}
\affiliation{Department of Chemistry, Princeton University, Princeton, NJ 08544, USA}
\author{H.-H. Klauss}
\affiliation{Institut f\"ur Festk\"orperphysik, Technische Universit\"at Dresden, 01062 Dresden, Germany}
\date{\today}
\begin{abstract}
We present results of $^{23}$Na and $^{19}$F nuclear magnetic resonance (NMR) measurements on NaCaCo$_2$F$_7$, a frustrated pyrochlore magnet with a Curie-Weiss temperature, $\Tcw \approx -140$\,K, and intrinsic bond disorder.
Below $3.6$\,K both the $^{23}$Na and $^{19}$F spectra broaden substantially in comparison to higher temperatures accompanied by a considerable reduction (80 \%) of the NMR signal intensity: This proves a broad quasi-static field distribution. 
The $^{19}$F spin-lattice relaxation rate $^{19}(1/T_1$) exhibits a peak at 2.9\,K already starting to develop below 10\,K.
We attribute the spin freezing to the presence of bond disorder. This is corroborated by large-scale Monte-Carlo simulations of a classical bond-disordered XY model on the pyrochlore lattice.
The low freezing temperature, together with the very short magnetic correlation length not captured by the simulations, suggesting that quantum effects play a decisive role in NaCaCo$_2$F$_7$.
\end{abstract}

\maketitle

The pyrochlore lattice is one of the canonical lattices exhibiting geometric frustration. Pyrochlore oxides with the general formula R$_2$T$_2$O$_7$ (R\,=\,rare earth ion, T\,=\,Ti, Sn, Mo, Ir etc.) are characterized by strong frustration of the rare-earth magnetic moments and have been found to display a variety of fascinating low-temperature phases, including classical and quantum spin-ice regimes, quantum order-by-disorder, and exotic spin liquids~\cite{chalker98,RevModPhys.82.53,Lacroix-Mendels-Mila}.

{\nccf} is a recently synthesized A$_2$B$_2$X$_7$-type pyrochlore \cite{PhysRevB.89.214401}. Here, the B site hosts high-spin Co$^{2+}$ in CoF$_6$ octahedra, whereas the A site contains a random distribution of Na and Ca ions, giving rise to exchange (bond) disorder for the magnetic degrees of freedom.
Bond disorder can lift degeneracies and lead to new unconventional ground states, both long-range ordered~\cite{zhito_prl13,maryasin14} and glassy~\cite{chalker_prl07,chalker10,motome_prl11,PhysRevB.89.054433-Y2Mo2O7}.

The uniform susceptibility of {\nccf} displays a Curie-Weiss law with a characteristic temperature $\Tcw \approx -140$\,K and a moment of $6.1\,\mub$ per Co, the latter suggesting an orbital contribution in addition to the $S=3/2$ state of Co$^{2+}$. Despite the  $\Tcw \approx -140$\,K, the material does not display a long-range-ordered magnetic state down to temperatures of $0.6$\,K ~\cite{PhysRevB.89.214401}. Instead, ac and dc susceptibility data indicate a spin freezing at $\Tf \approx 2.4$\,K~\cite{PhysRevB.89.214401}, yielding a frustration index of $f=\Tcw/\Tf\approx56$. The observed entropy loss at the freezing transition is low, suggesting that magnetic entropy remains present at least down to $0.6$\,K.
Inelastic neutron scattering (INS) results \cite{Kate-Ross2015} for {\nccf} show diffuse elastic scattering below $2.4$\,K.  
The momentum-space structure of the spin correlations below $2.5$\,meV has been interpreted in terms of XY-like antiferromagnetic clusters, indicating a local easy-plane anisotropy at low energies, whereas a signal corresponding to collinear correlations was found above this energy. The XY character together with the short detected correlation length of $16$\,{\AA} \cite{Kate-Ross2015} appears to be at odds with the theoretical suggestion \cite{maryasin14} that bond disorder in an XY pyrochlore magnet induces long-range magnetic order of the so-called $\psi_3$ type (as opposed to $\psi_2$ order, which is selected by quantum or thermal fluctuations \cite{zhitomirsky12,savary12,zhitomirsky14}). More recently, NaSrCo$_2$F$_7$ system is reported which exhibits similar static and dynamic correlation~\cite{{PhysRevB.95.144414},{ross-2017}}.
Hence, the interplay of frustration and quenched disorder in {\nccf} prompts further studies. In addition, given that there are only few reports available in the literature covering the NMR studies on frustrated pyrochlore system, the present NMR investigations on {\nccf}\ are by itself interesting.

Here we present a detailed NMR investigation of {\nccf}, thereby microscopically accessing the magnetic order and the low-energy spin dynamics on a $\mu$eV energy scale. We characterize the short-range-ordered state at low temperature: It displays substantial NMR line broadening in comparison to high temperature and a wipeout effect characteristic of glassy spin freezing. A peak of the spin-lattice relaxation rate at low temperature is well described by thermally activated low-energy spin fluctuations.

We complement these experimental results by large-scale Monte-Carlo simulations of a classical pyrochlore XY model with bond disorder. The simulation results show that, indeed, bond disorder can lead to a glass-like state without long-range order, qualitatively consistent with experiments.
However, quantitative comparison between experiment and theory shows disagreement, implying that physics beyond the classical XY model is important in {\nccf}.
We discuss the possibility that the material is a bond-disordered quantum spin liquid.


\section{Measurements}
\begin{figure}
\includegraphics[width=\columnwidth]{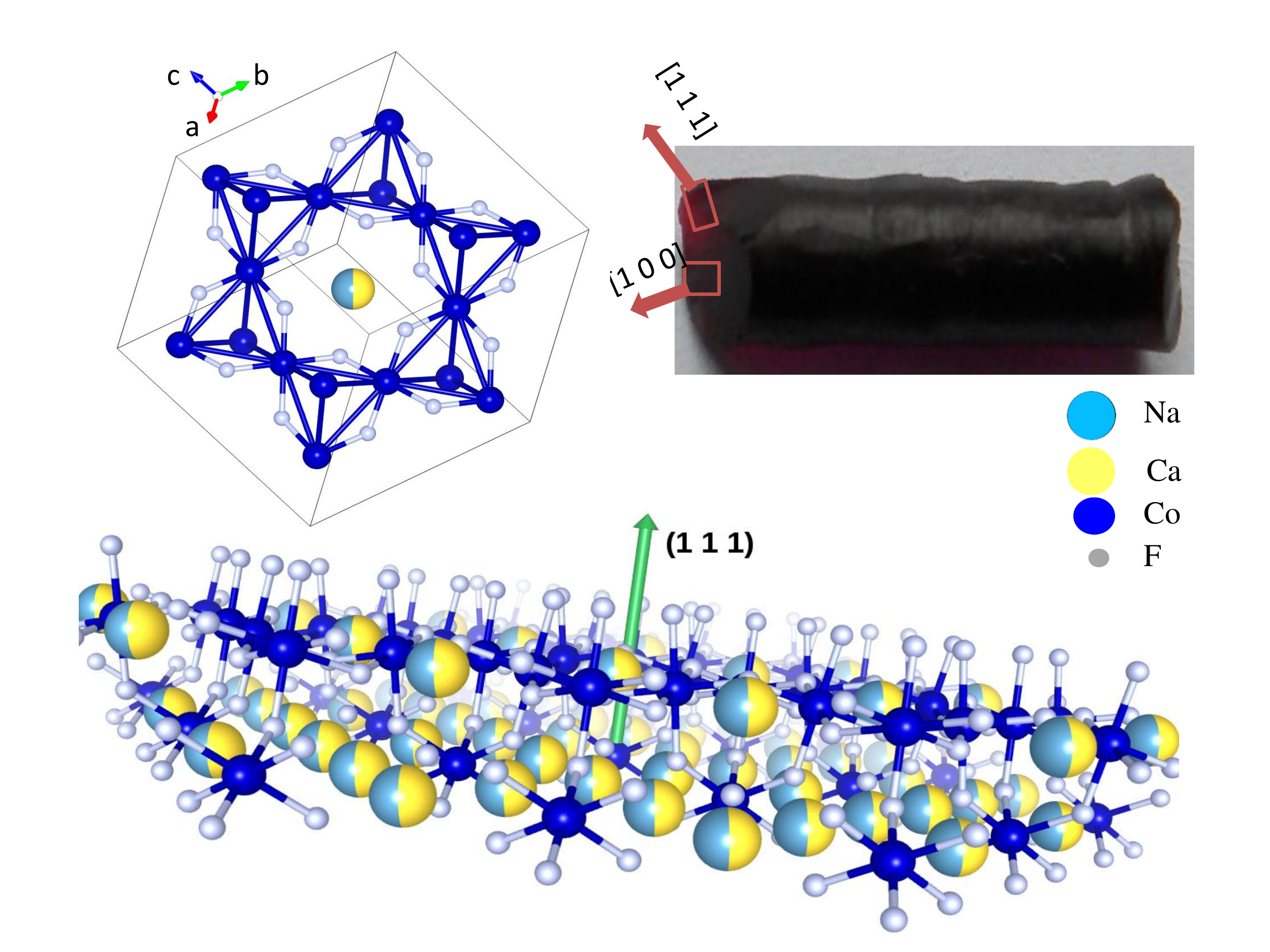}
\caption{\label{fig:fig1} Top left panel represents the crystallographic view along the [1 1 1] direction of NaCaCo$_2$F$_7$. Lower panel depicts the side view of the (Na or Ca) layers. The top right part of the figure shows a picture of the single crystal employed.
%
}\label{fig1}
\end{figure}

%
Single crystals of {\nccf} were prepared in an optical floating zone furnace, see  Ref.~\onlinecite{PhysRevB.89.214401} for details. NMR experiments were performed using a Tecmag spectrometer at different frequencies and in a wide temperature range. A crystal of mass 0.89\,g was measured, with fields applied along the [1 0 0], [1 1 0], and [1 1 1] directions, see Figure~\ref{fig1}.
We conducted the $^{23}$Na (I=3/2, $\gamma$\,=\,11.2625\,MHz/T, 100\% abundance) and $^{19}$F (I=1/2, $\gamma$\,=\,40.0594\,MHz/T, 100\% abundance) line shape measurements using a 90$^{\circ}$-90$^{\circ}$ pulse sequence. For spin-lattice relaxation 1/$T_1$ measurements we used the saturation recovery method with a 90$^{\circ}$-90$^{\circ}$ pulse sequence. 1/$T_1$ experiments were performed by exciting the central position of the spectra.

\section{\label{sec:level1}NMR}

\subsection{$^{23}$Na NMR spectra} 

\begin{figure}
\includegraphics[width=\columnwidth]{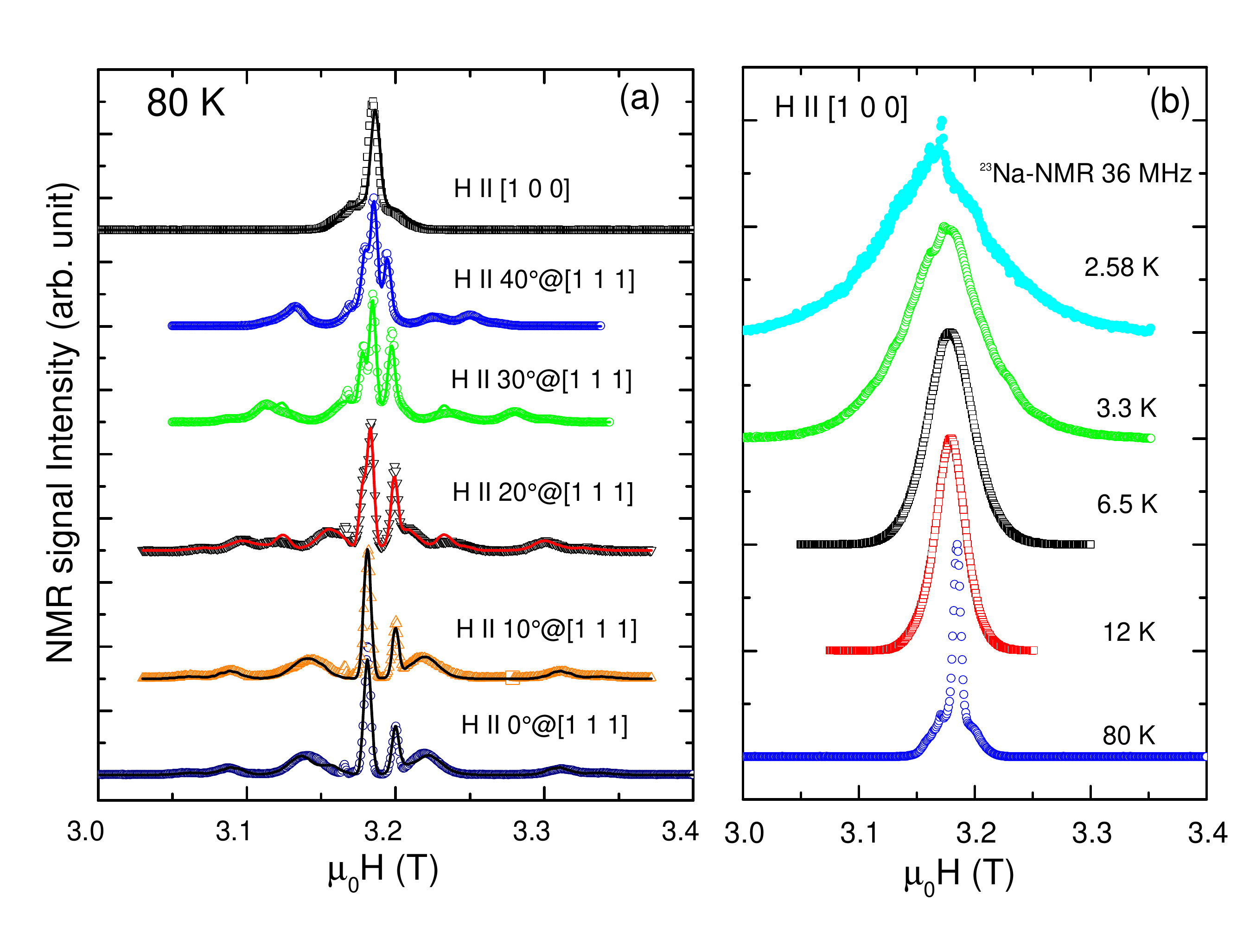}
\caption{\label{fig:fig2}a)Angular dependence of the $^{23}$Na NMR field-sweep spectra from [1 1 1] to [1 0 0] direction at 80\,K at the frequency of 36\,MHz. Lines represent the theoretical descriptions. For details see the main text. b) $^{23}$Na NMR field-sweep spectra at the frequency of 36\,MHz for a field along the [1 0 0] direction.}
\label{fig2}
\end{figure}

Figure~\ref{fig2}(a,b) shows the angle dependent, with respect to the external magnetic field, $^{23}$Na NMR field-sweep spectra from [1 1 1] to [1 0 0] direction at 80\,K and at representative temperatures for [1 0 0] direction, respectively. The spectra can be described by diagonalizing the full NMR static Hamiltonian. However, for the complete description of the experimental data we need to take into account the (charge and bond) disorder effect of the A site (Na$^{+}$/Ca$^{2+}$) of NaCaCo$_2$F$_7$(for details see below). 
As the magnetic field applied is along the [1 1 1] direction the appearance of two Na lines indicates that Na nuclei locally experience two different hyperfine fields. This is the microscopic reason for this double-peak structure in the [1 1 1] direction. Similar features were also reported in another pyrochlore by means of Ti NMR although Ti nucleus is located at the B site of A$_2$B$_2$O$_7$ type pyrochlore ~\cite{PhysRevB.77.214403}.

The $^{23}$Na spectra broaden gradually upon cooling, 
resulting in a very broad line shape in comparison with the spectra at high temperatures. $^{23}$Na spectra broaden substantially at the same temperature range where ac susceptibility data show a broad maximum \cite{PhysRevB.89.214401}. Such a spectrum proves a quasi-static and spatially disordered field distribution in {\nccf}. Collinear and non-collinear long-range ordered states can be ruled out, because in single crystals those would induce a distinct line splitting in the ordered state~\cite{{PhysRevB.64.024409},{PhysRevB.84.214403},{doi:10.1143/JPSJ.77.114709}}.

\subsection{Analysis of the $^{23}$Na NMR spectra}

\begin{figure}[h]
\begin{center}
\includegraphics[width=0.5\textwidth]{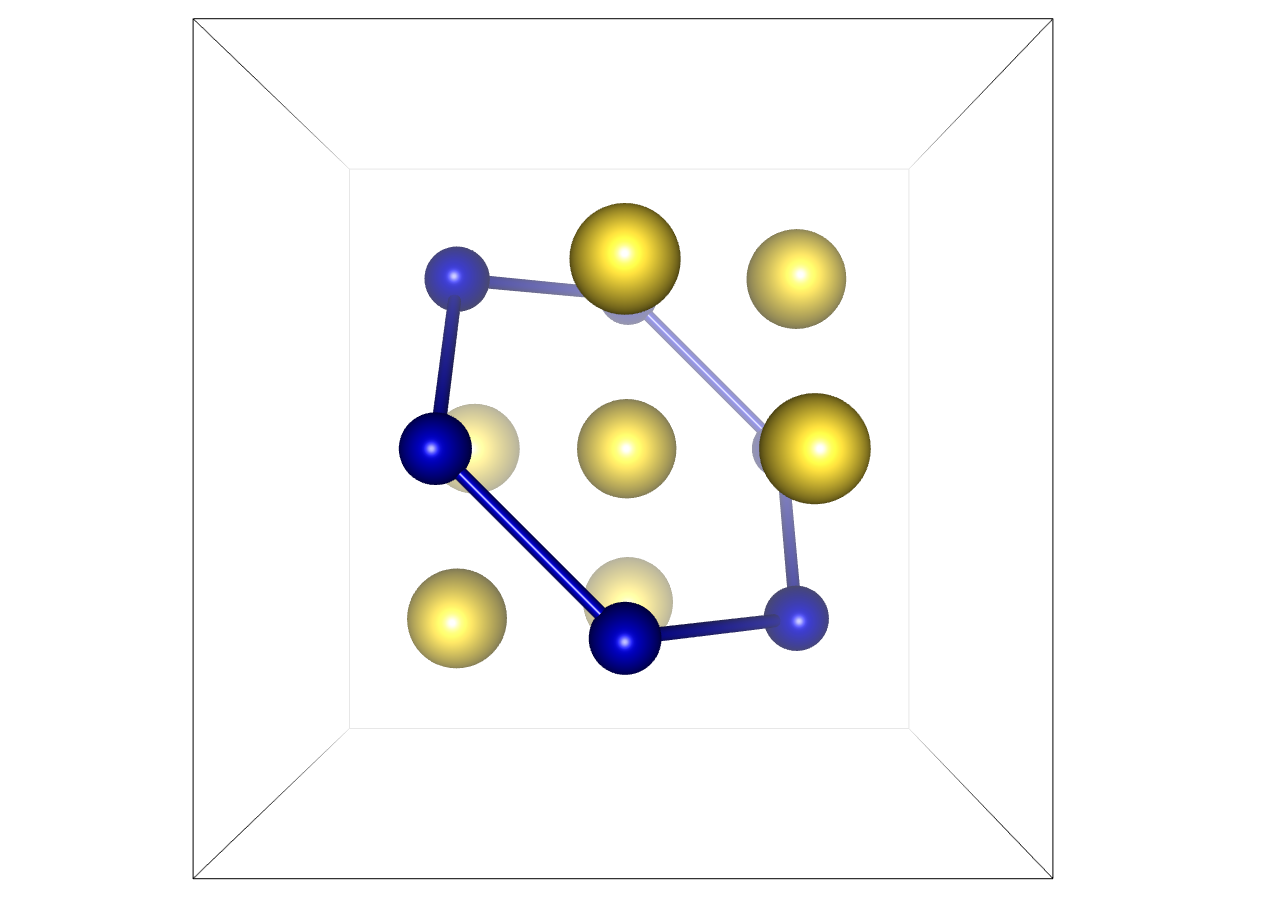}
\caption{Local environment of a Na atom (central yellow ball). Yellow balls depict Na/Ca and blue balls depict Co atoms. Fluorine is not shown.}\label{uc}
\end{center}
\end{figure}

$^{23}$Na NMR spectra were taken on the geodesic between $(1\,1\,1)$ and $(1\,0\,0)$ directions. To describe the data, a physical model is constructed, by taking into account the crystal structure and the nearest neighbors environments. We describe the NMR spectra by diagonalizing the full spin Hamiltonian without further approximations.
The NMR spin Hamiltonian consists of the Zeeman term and the quadrupolar term. The local magnetic hyperfine field is the extrinsic parameter in the Zeeman term, so is the electric field gradient (EFG) in the quadrupolar term. To make assumptions about the extrinsic parameters in the general Hamiltonian, a deeper look into the structural environment of Na is important: In Fig. \ref{uc}, the local environment of a Na atom is shown. Now we start with a uniform pyrochlore lattice. Because of the $\bar{3}$ rotation symmetry of space group $227$ in hkl [1 1 1] direction, the EFG necessarily has a defined eigen-vector hkl [1 1 1] and two degenerate eigen-vectors perpendicular to hkl [1 1 1]. The so-called asymmetry parameter $\eta$ is zero. 
\\
When introducing the charge disorder, charges of $0.5\,e$ and $-0.5\,e$ distribute on the Na/Ca site. So the deviation of the EFG can be calculated with a nearest neighbor point charge model, leading to $2^6 = 64$ different Na environments and $3^3 = 27$ different EFG's. Due to the unknown polarization of individual magnetic Co ions, the local hyperfine field is approximated to be a phenomenological cosine-function of the angle between hkl [1 1 1] and the external magnetic field, motivated by the ring-like structure of nearest neighbor Co sites (see Fig. \ref{uc}). The Knight shift parallel to hkl [1 1 1] direction is found to be $K_\parallel = - 0.0011$ and perpendicular $K_\perp = 0.0057$. When reviewing the data one can determine a satellite double peak structure in the hkl [1 1 1] direction. Therefore we assume two sub-spectra with nuclear quadrupolar interaction parameter $\nu_{q,1} = 1.165\,\mathrm{MHz}$ and $\nu_{q,2} = 1.480\,\mathrm{MHz}$ with $\approx 80\%$ respectively $\approx 20\%$ intensity. All other parameters are the same for each sub-spectrum. Furthermore, we account for a $1.5^\circ$ tilting of the crystal. One further fitting parameter is the ratio of satellite intensity to the central line intensity. This has to be a free parameter due to the unknown quadrupolar relaxation and the transition-dependent matrix elements of the $B_1$ operator (ladder operators) in a spin 3/2 nucleus. The Hamiltonian is now diagonalized for hkl [1 1 1], hkl[-1 1 1], hkl[1 -1 1] and hkl[1 1 -1] main axis directions of the EFG for all nearest neighbors environments, both sub-spectra and the crystal orientations used in experiments.
\\
\indent The calculation of the field sweep spectrum assuming a narrow gaussian NMR excitation gives sufficient fits to the data. Therefore one can state the charge disorder being the driving force of the quadrupolar broadening and estimate the local lattice distortion to be weak ($< 1\%$). The two sub-spectra with $\nu_{q,1} = 1.165\,\mathrm{MHz}$ and $\nu_{q,2} = 1.480\,\mathrm{MHz}$ can be interpreted as the breakdown of the point charge model: For some nearest neighbor configurations, the point charge model underestimates the impact on the EFG. Parenthetically we note that the $\nu_Q$ as obtained from the fits are close to the values as found in other pyrochlores.~\cite{{Kobayashi2001347},{PhysRevB.77.214403}}


\section{$^{19}$F NMR spectra} 
\begin{figure}[h]
\includegraphics[width=\columnwidth]{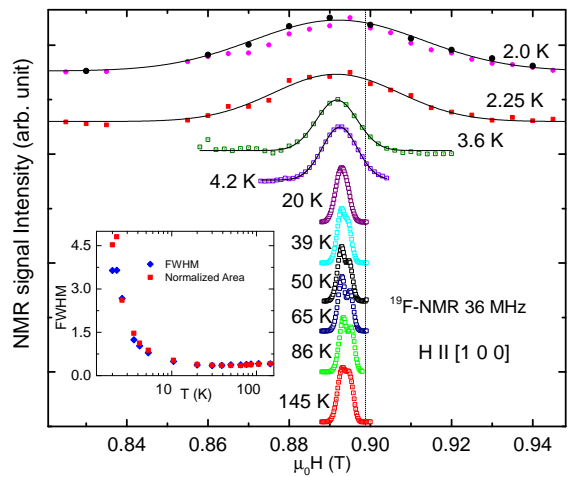}
\caption{\label{fig:fig4}$^{19}$F NMR field-sweep spectra at the frequency of 36\,MHz for a field along the [1 0 0] direction. Lines indicate Gaussian single-peak fit. Inset shows the FWHM and the normalized area as a function of temperature. The
spectra are normalized to the maximum (peak value) of NMR signal. Vertical dotted line represents the position of the Larmor field (diamagnetic reference). }
\label{fig4}
\end{figure}

In Fig. \ref{fig4} we show the $^{19}$F field-sweep NMR spectra at representative temperatures. The $^{19}$F spectra are shifted to lower magnetic fields in comparison with the $^{19}$F bare nuclei Larmor frequency. Below $\sim3.6$\,K the spectra broaden similar to $^{23}$Na spectra. The effective line width at the FWHM position increases by a factor of 16 when temperature is lowered from 145\,K to 2\,K. Such a broadening proves the presence of a broad distribution of static internal fields. At high temperature a fine structure of the $^{19}$F spectra is observed which is associated with two inequivalent F1 and F2 positions in this pyrochlore.

\begin{figure}
\includegraphics[width=\columnwidth]{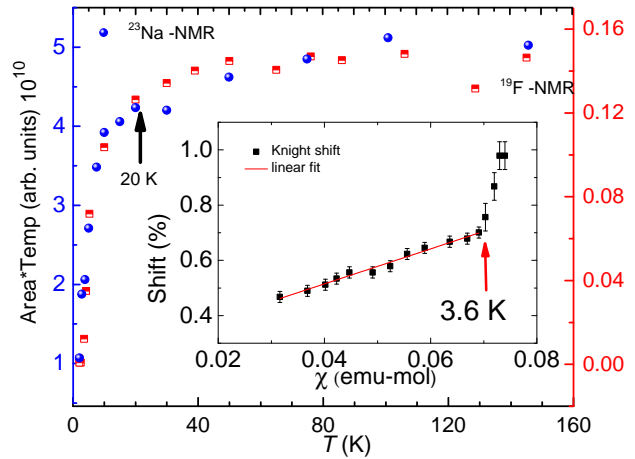}
\caption{\label{fig:fig5} The wipe-out effect of $^{23}$Na and $^{19}$F NMR signal. 
Inset represents the plot of shift vs. susceptibility. The line indicates the linear fit.
}\label{fig5}
\end{figure}

The NMR spin-echo intensities multiplied with the temperature are plotted as a function of temperature in Fig.~\ref{fig5} to have a clear picture of this field distribution.
The total NMR signal intensity, i.e. the area under the curve, is in general inversely proportional to the temperature and proportional to the number of contributing nuclei. The strong loss of signal intensity with lowering the temperature proves that at low temperature not all nuclei contribute to the observed NMR signal. A broad distribution of spin-lattice relaxation rates may lead to such a situation ~\cite{{PhysRevLett.85.642},{PhysRevB.64.134525},{PhysRevLett.82.4300},{Mendels-PhysRevLett.85.3496}}. In the present case a robust loss of the $^{19}$F and $^{23}$Na NMR signal intensities is observed: The signals start to decrease drastically below 20\,K, in the same temperature regime where $^{19}(1/T_1)$ starts to increase, see Fig.\,\ref{fig6}. Hence, the development of slow spin fluctuations and the loss of signal intensity are correlated.

The reduction of the NMR signal intensity, the so-called "wipeout effect", has been widely discussed in the literature in the context of spin-frozen states, and {\nccf} fits into this picture \cite{{ PhysRevLett.82.4300},{PhysRevLett.85.642},{MacLaughlin-spinglass-1976}}.

The inset of Fig.~\ref{fig5} plots the NMR shift as a function of the bulk susceptibility. This dependence is linear down to $3.6$\,K, but deviates for lower temperatures. To interpret this we recall that the total shift is the sum of the chemical shift $K_0$ and the Knight shift $K$: While $K_0$ represents the $T$-independent orbital and conduction-electron contribution, $K=A\chi(T)$ describes the hyperfine coupling to the electronic moments mainly residing on the Co atoms, with $A$ being the hyperfine coupling constant and $\chi(T)$ the  magnetic susceptibility. The linear dependence above $3.6$\,K yields an estimate for the hyperfine coupling of $A=0.354$~kOe$/\mu_B$. The change of slope at $3.6$\,K reflects an increase of the hyperfine coupling constant and indicates a change of the electronic state from correlated paramagnetic to quasi-static ordered. 

Such behavior is often observed, e.g., at the formation of correlated electronic states in heavy-fermion systems~\cite{PhysRevB.90.241109-curro}. Whereas, in Ref.[\onlinecite{PhysRevB.90.241109-curro}] the change in $A$ is small, the change of $A$ in the NaCaCo$_2$F$_7$ data is larger. This situation is more likely associated with a phase transition.

\begin{figure}[b]
\includegraphics[width=\columnwidth]{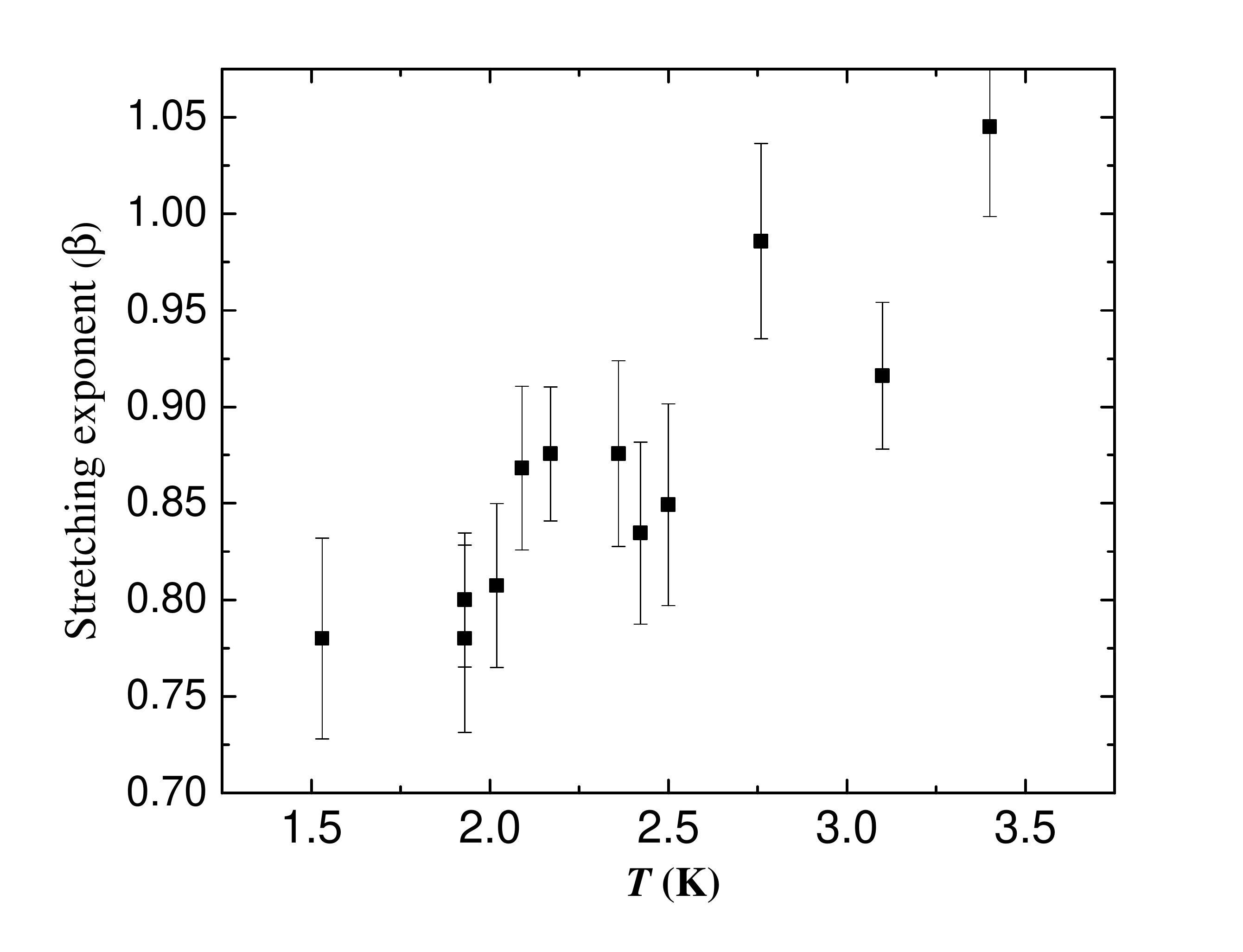}
\caption{\label{fig:spl1} Stretching exponent $\beta$ as function of temperature. This $\beta$ is obtained by using equation \ref{eq:equation1} to fit the $^{19}$F spin-lattice recovery data. 
}
\end{figure}

\begin{figure}
\includegraphics[width=\columnwidth]{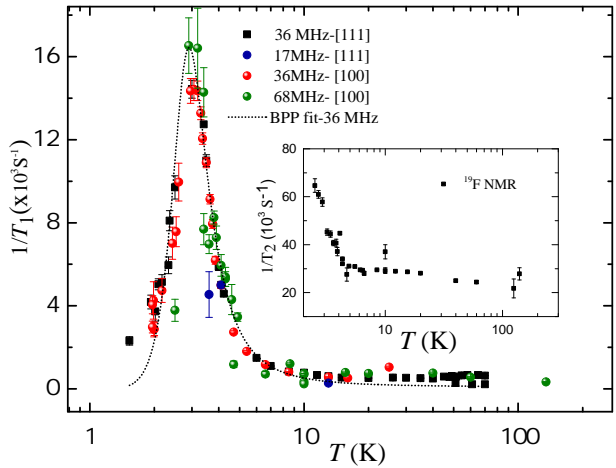}
\caption{\label{fig:fig6}$^{19}$F spin-lattice relaxation rate at different fields and different crystallographic orientations.  Inset shows the temperature dependence of $^{19}(1/T_2)$.}\label{fig6}
\end{figure}

\section{Spin relaxation}
Next, we discuss the spin dynamics at very low energies. 
The $^{19}$F spin-lattice recovery data were well described by the following equation:
\begin{equation}
\label{eq:equation1}
m^{\prime}(t)= \frac{m^{\prime}_0 - m^{\prime}(t)}{m^{\prime}_0}~=A\exp[-(t/T_1)^\beta],
\end{equation}
where $m^{\prime}_0$ is the equilibrium nuclear magnetization and $\beta$ is the stretching exponent.
 For a system with homogeneous and unique relaxation rate, a fit to the spin-lattice recovery curve in principle gives $\beta=1$. In contrast, a best fit with $\beta<1$ indicates a distribution of relaxation rates. This is observed in spin-frozen systems which exhibits dynamical inhomogeneity. In the present case, we observed $\beta=1$ in the temperature range $4-150$\,K, however below $4$\,K the fit gave $\beta<1$.
In Fig.~\ref{fig:spl1} we have shown the $\beta<1$ values as a function of temperature below 4\,K.

In Fig.~\ref{fig6} we show the $^{19}$F NMR spin-lattice relaxation rate at different fields as a function of temperature. $^{19}$(1/$T_1$) values are constant in the temperature range $10-150$\,K, however below $10$\,K, $^{19}$(1/$T_1$) starts to increase exhibiting a peak at around $2.9$\,K. Ac and dc susceptibility data also show a broad maximum around this temperature. The spin-spin relaxation rate $^{19}$(1/$T_2$) has  a similar temperature dependence as $^{19}$(1/$T_1$), although $^{19}$(1/$T_2$) values are much larger than $^{19}$(1/$T_1$).
The enhancement of $^{19}$(1/$T_1$) below $10$\,K suggests that in {\nccf} a fluctuating correlated magnetic state is formed, i.e., a cooperative paramagnet, before it settles into a frozen state. These findings are in good agreement with neutron scattering data~\cite{Kate-Ross2015}. Nevertheless, it is worth mentioning in dilute paramagnets the similar temperature dependencies of the nuclear spin-lattice relaxation rates could also be observed. Note that the $^{19}$(1/$T_1$) relaxation function become "stretched" below 4\,K. The stretched exponent ($\beta$) values less than one indicate the development of disorder in {\nccf}.

In Fig.~\ref{fig6}, above 10\,K, the $1/T_1$ data appear to be
temperature independent. However, it is clearly noticeable that the spread of the points or the error bars are huge, more than a factor of two. This is somewhat hidden by the usage of the linear scale because the absolute values
are small, but it is clear that showing the data on a logarithmic scale would reveal the problem.

The NMR experiments do not show a strong crystallographic orientation dependence of $^{19}$(1/$T_1$) in the proximity of the peak, indicating that the spin fluctuations are macroscopically isotropic in nature, consistent with a randomly frozen spin state.

In the past \cite{{PhysRevLett.85.642},{1367-2630-11-7-075004},{PhysRevB.61.R9265}} certain classes of spin-frozen systems have been found to display a pronounced low-temperature peak in $1/T_1$ that can be described by the mechanism due to Bloembergen, Purcell, and Pound (BPP)~\cite{PhysRev.73.679}. This assumes a fluctuating hyperfine field $h(t)$ at the nuclear site with autocorrelation function $\langle h(t)h(0) \rangle = h_0^2 \exp(-t/\tau_c)$, where $\tau_c$ is the correlation time. In spin-frozen systems, $\tau_c$ often exhibits an activated ($\tau_c=\tau_0\exp(E_a/T$)) or Vogel-Fulcher ($\tau_c=\tau_0\exp(E_a/(T-T_s)$)) behavior\cite{PhysRevLett.85.836}. We have chosen 
activated behavior to describe the present data. In the presence of fluctuating hyperfine fields, $1/T_1$ as measured by NMR can be represented by the following equation in the framework of the BPP theory:
\begin{equation}
\label{eq:spectra-simulation}
1/T_1(T)=\gamma^2 h_0^2\tau_c(T)/[1 +\omega_L^2\tau_c^2(T)],
\end{equation}
where $\gamma$ is the nuclear gyromagnetic ratio, and $\omega_L$ is the Larmor frequency. This equation yields a peak in $1/T_1$ as function of $T$ once $1/\tau_c$ matches the measured resonance frequency. In Fig.~\ref{fig6} a dotted line describes the BPP model. Our fit yields, in case of 36\,MHz, $E_a=1.378\times10^{-3}$\,eV, $\tau_0=8.052\times10^{-11}$\,s, and $h_0=42$\,G. These values are in line with ac susceptibility experiments~\cite{PhysRevB.89.214401} and fit in the framework of spin glasses or spin-glass-like materials~\cite{0022-3719-18-17-001-spinglass}. Surprisingly, no significant frequency dependence of the $^{19}$(1/$T_1$) is observed (see Fig.~\ref{fig6}). The absence of the frequency dependence, in this frequency range, points to effects beyond the BPP description, possibly related to unconventional disorder.


\section{Monte-Carlo simulations}
In order to study the qualitative question whether bond disorder can induce glassy behavior, we have performed finite-$T$ Monte-Carlo simulations of a classical pyrochlore XY model with and without bond disorder on lattices with linear size up to $L\,=\,12$. In line with earlier work, the clean system orders in the $\psi_2$ state due to a thermal order-by-disorder effect \cite{zhitomirsky12,savary12}.
Upon introducing strong bond disorder, $\psi_2$ and $\psi_3$ configurations are expected to compete \cite{maryasin14}. This is analyzed by monitoring the probability distribution of the quantity
\begin{equation}
m_{6}=m\cos (6\theta),\label{m6eq}
\end{equation}
where the angle $\theta=\tan^{-1}\left(m_{y}/m_{x}\right)$ describes the in-plane spin direction in a local coordinate frame, such that $m_{6}=1$ ($-1$) holds for an ideal $\psi_{2}$ ($\psi_3$) state, respectively. 
Here we introduced the magnetic order parameter $m=\sqrt{ m_x^2 + m_y^2}$.
Sample results for the $m_6$ histogram are shown in Fig.~\ref{figmc}, with a clear bimodal distribution below the freezing temperature, for details see Appendix. We conclude that a long-range ordered $\psi_3$ state is not realized, at variance with the results of Ref.~\onlinecite{maryasin14}. Instead, bond disorder causes the system to settle into a glassy phase with a spatial mixture of $\psi_2$ and $\psi_3$ configurations. This appears qualitatively consistent with the experimental data on {\nccf} and Ref.~\onlinecite{{note-33}, {note-34}}.

\begin{figure}
\includegraphics[width=0.6\columnwidth]{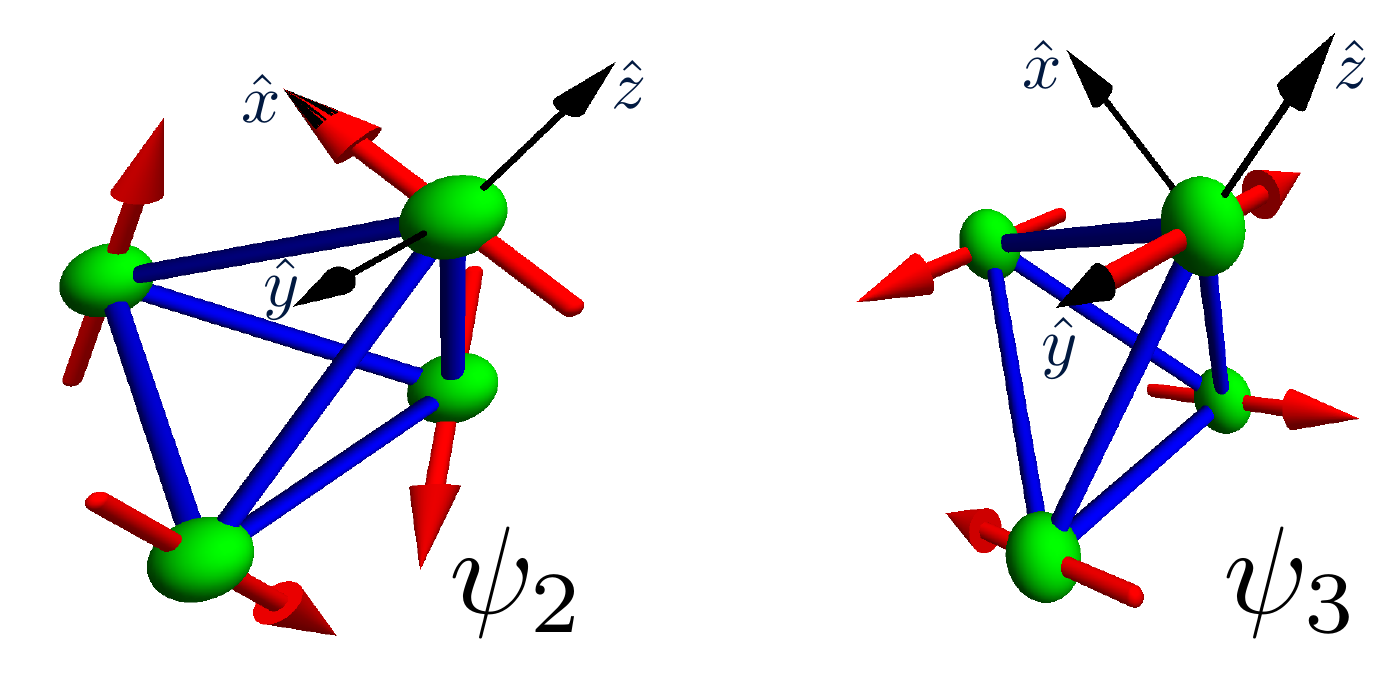}
\includegraphics[scale=0.3]{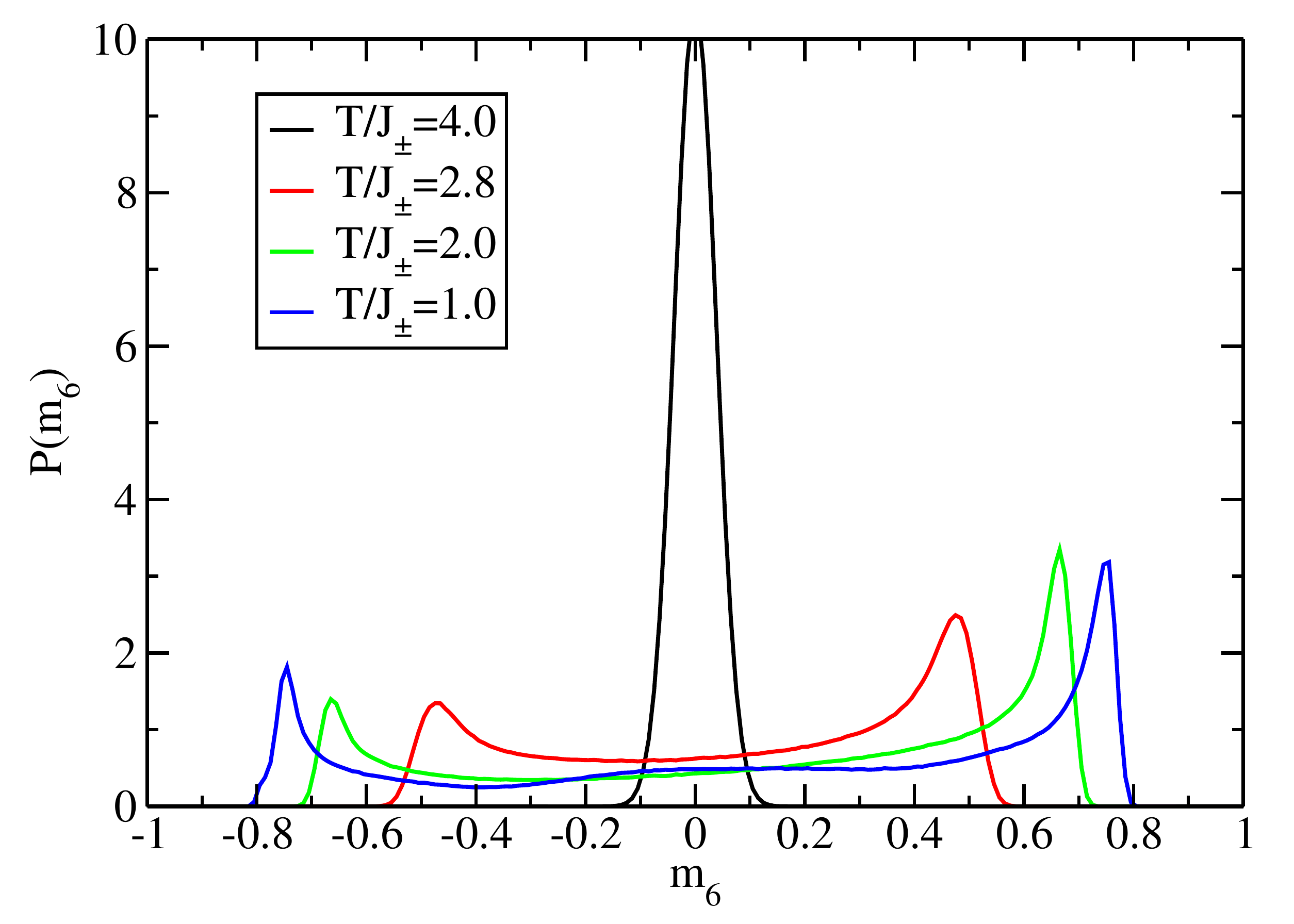}
\caption{
Top: Schematic representation of the $\psi_2$ and $\psi_3$ states of pyrochlore XY magnets. The orthogonal local axes $\hat{x}_{\alpha}\perp\hat{y}_{\alpha}$ on each site of a tetrahedron coincide with spin directions for the two states.
Bottom: Histogram of $m_6$ (\ref{m6eq}) -- distinguishing $\psi_2$ and $\psi_3$ --  for the disordered pyrochlore XY model obtained from MC simulations at different temperatures; the freezing temperature is $\Tf/J_{\pm}=3.15\left(2\right)$. The bond strengths are taken from a box distribution with width $W/J_{\pm}$=2, where $J_{\pm}$ is the exchange integral; for details see Appendix.
\label{figmc}
}
\end{figure}

However, a quantitative comparison shows that our simulations overestimate both the magnetic correlation length and the freezing temperature even for strong bond disorder (see Appendix). Hence, fluctuation effects must be crucial, with two obvious sources, namely quantum effects and fluctuations out of the XY manifold of states due to small crystal field splitting.
Theoretical investigations of ``clean'' spin models on the pyrochlore lattice have shown that a variety of ordered states compete, and that quantum fluctuations are likely to stabilize quantum spin liquids in parameter regimes near classical phase boundaries \cite{savary12b,PhysRevB.95.094422,gingras_rop14}. Given the large frustration ratio $\Tcw/\Tf$ of {\nccf}, we speculate that it may be understood as a quantum spin liquid which is driven into a weakly glassy state by bond disorder.

\section{Conclusions}
Using NMR measurements we have characterized the magnetism of {\nccf}. It displays (i) a huge reduction of the NMR signal intensity at low temperature, (ii) a broad distribution of static internal field at low temperature, and (iii) a peak in the $^{19}(1/T_1)$\,vs.\,$T$ data which can be explained by the thermally activated autocorrelation function.
Hence, {\nccf} represents a frustrated pyrochlore material with unusual static local antiferromagnetic correlations which stabilize a spin-frozen state. 
Via Monte-Carlo simulations we have qualitatively shown that the glassy freezing can be attributed to quenched structural disorder which in turn induces magnetic bond disorder. The presence of strong low-energy spin fluctuations, as evidenced by neutron scattering~\cite{Kate-Ross2015} and NMR, indicates that quantum effects are crucial for a quantitative understanding of the material.


\section{ACKNOWLEDGMENTS} We acknowledge helpful discussions with K. A. Ross and H. Yasuoka. This research is supported by the Deutsche Forschungsgemeinschaft (DFG) through SFB 1143. The crystal growth work at Princeton University was supported by the US DOE Division of Basic Energy Sciences, grant DE-FG02-08ER46544. E.C.A. was supported by FAPESP (Brazil) Grant No. 2013/00681-8.

\appendix
\section{Monte-Carlo simulations}

In this section we describe our Monte-Carlo simulations for a disordered XY model on the pyrochlore lattice, together with theoretical background. We recall that the main purpose of these simulations is to answer the qualitative question whether or not spin freezing (instead of long-range order) can be induced by bond disorder in a pyrochlore XY antiferromagnet.

\subsection{Pyrochlore XY antiferromagnet}

The pyrochlore antiferromagnet with local XY anisotropy (see below for a precise definition) is an interesting example of frustrated magnetism. At the classical level, it features a one-parameter manifold of degenerate states with ordering wavevector $\vec{Q}=0$. Special states in this manifold are the non-coplanar $\psi_{2}$ state and the coplanar $\psi_{3}$ state; see Fig. \ref{figmc} (top panel). The two states form a basis of the $E$ irreducible representation of the tetrahedral point group and can be continuously turned into
each other by simultaneous rotation of four sublattices. The degeneracy between $\psi_{2}$ and $\psi_{3}$ persists even with an extra biquadratic exchange, however, both quantum and thermal fluctuations choose the noncoplanar $\psi_{2}$ state,\citep{zhitomirsky12,savary12,champion04,zhitomirsky14,mcClarty14,javanparast15}
an instance of the phenomenon known as order by disorder.\citep{villain80}
An experimental example is provided by {\ertio}. This pyrochlore material undergoes a second-order transition at $T_{N}=1.2$\,K into the $\psi_{2}$ state-\citep{champion03} In fact, in the context of {\ertio} it has been suggested that quenched disorder, either in form of vacancies or bond disorder, may induce long-range order of the coplanar $\psi_{3}$ type instead.\citep{maryasin14,andreanov15}
However, the combination of frustration and quenched disorder can be generically expected to lead to spin-glass physics, which prompts us to investigate the pyrochlore XY model with disorder in more detail.


\subsection{Effective spin Hamiltonian}

In cubic pyrochlore materials, magnetic ions form a network of corner-sharing
tetrahedra. The unit cell contains four magnetic sites, with positions
\begin{align}
&\vec{r}_{0}=\left(0,0,0\right),\,&\vec{r}_{1}=\left(\frac{1}{4},\frac{1}{4},0\right),\notag\\
&\vec{r}_{2}=\left(0,\frac{1}{4},\frac{1}{4}\right),\,&\vec{r}_{3}=\left(\frac{1}{4},0,\frac{1}{4}\right),
\label{eq:pyro_basis}
\end{align}
and the primitive lattice vectors are chosen as those of a fcc lattice
\begin{equation}
\vec{a}_{1}=\left(\frac{1}{2},\frac{1}{2},0\right),\,\vec{a}_{2}=\left(0,\frac{1}{2},\frac{1}{2}\right),\,\vec{a}_{3}=\left(\frac{1}{2},0,\frac{1}{2}\right),\label{eq:primitive_fcc}
\end{equation}
where the cubic lattice parameter has been set to unity.
The directions of local $\left\langle 111\right\rangle $ axes, connecting the centers of neighboring tetrahedra, are given by
\begin{align}
&\hat{z}_{0}=\frac{1}{\sqrt{3}}\left(1,1,1\right),\,&\hat{z}_{1}=\frac{1}{\sqrt{3}}\left(-1,-1,1\right),\notag\\
&\hat{z}_{2}=\frac{1}{\sqrt{3}}\left(1,-1,-1\right),\,&\hat{z}_{3}=\frac{1}{\sqrt{3}}\left(-1,1,-1\right).
\label{eq:local_z}
\end{align}

In the pyrochlore XY magnet the magnetic moments explore planes perpendicular to these local $\hat{z}$ axes. To describe the classical ground states we choose the directions of the $\hat{x}_{i}$ and $\hat{y}_{i}$ axes such that they coincide with the sublattice direction for the $\psi_{2}$ and the $\psi_{3}$ state, Fig. \ref{figmc} (top panel), respectively:
\begin{align}
&\hat{x}_{0}=\frac{1}{\sqrt{6}}\left(1,1,-2\right),\,~~~\hat{x}_{1}=\frac{1}{\sqrt{6}}\left(-1,-1,-2\right),\notag\\
&\hat{x}_{2}=\frac{1}{\sqrt{6}}\left(1,-1,2\right),\,~~~\hat{x}_{3}=\frac{1}{\sqrt{6}}\left(-1,1,2\right).
\label{eq:local_x}
\end{align}
and
\begin{align}
&\hat{y}_{0}=\frac{1}{\sqrt{2}}\left(-1,1,0\right),~~~\hat{y}_{1}=\frac{1}{\sqrt{2}}\left(1,-1,0\right),\notag\\
&\hat{y}_{2}=\frac{1}{\sqrt{2}}\left(-1,-1,0\right),\,~~~\hat{y}_{3}=\frac{1}{\sqrt{2}}\left(1,1,0\right).
\label{eq:local_y}
\end{align}

To build a spin model, we start by assuming that the crystalline electrical field, with trigonal symmetry, splits the energy levels of each magnetic ion into Kramers doublets. As a result, at temperatures below the crystal-field splitting each moment can be represented by an effective pseudospin $1/2$. 

The effective exchange Hamiltonian is a bilinear form of these pseudospin operators which is symmetric under corresponding crystal-lattice transformations. The most general nearest-neighbor exchange Hamiltonian allowed by symmetry reads \citep{maryasin14}
\begin{eqnarray}
\mathcal{H} & = & \sum_{\left\langle ij\right\rangle }\big\{ J_{zz}S_{i}^{z}S_{j}^{z}-J_{\pm}\left(S_{i}^{+}S_{j}^{-}+S_{i}^{-}S_{j}^{+}\right)\nonumber\\
&+&J_{\pm\pm}\left(e^{i\theta_{ij}}S_{i}^{+}S_{j}^{+}+e^{-i\theta_{ij}}S_{i}^{-}\cdot S_{j}^{-}\right)\nonumber \\
 & - & \left.J_{z\pm}\left[S_{j}^{z}\left(e^{-i\theta_{ij}}S_{i}^{+}+e^{i\theta_{ij}}S_{i}^{-}\right)+i\longleftrightarrow j\right]\right\}\label{eq:spin_pyro_gen}
\end{eqnarray}
where the spin components refer to {\em local} coordinate frames -- one for each member of the tetrahedral basis -- with the $z$ axes along the $\left[111\right]$ directions. This form of the spin Hamiltonian was previously employed in a number of works \citep{savary12,ross11,wong13} with a minor redefinition of complex factors. For our particular choice of the local frame we get
\begin{equation}
\begin{array}{ccc}
\theta_{01}=\theta_{23}=0, & \theta_{02}=\theta_{13}=2\pi/3, & \theta_{03}=\theta_{12}=-2\pi/3\end{array}.\label{eq:phases}
\end{equation}

In the following, we take $J_{zz}=J_{z\pm}=0$ so that we are left with a nearest-neighbor XY model; recall that the local easy planes are globally noncoplanar. This model has a single tuning parameter $J_{\pm\pm}/J_{\pm}$. Writing $\vec{S}_{i}=\left(S_{i}^{x},\, S_{i}^{y},\,0\right)$ in the local frame and utilizing $S_{i}^{\pm}=S_{i}^{x}\pm iS_{i}^{y}$ this XY model reads
\begin{equation}
\mathcal{H}  = \sum_{\left\langle ij\right\rangle }\left\{ J_{ij}^{xx}S_{i}^{x}S_{j}^{x}+J_{ij}^{yy}S_{i}^{y}S_{j}^{y}+J_{ij}^{xy}\left(S_{i}^{x}S_{j}^{y}+S_{i}^{y}S_{j}^{x}\right)\right\}
\label{eq:spin_pyro_xy_aniso}
\end{equation}
where we defined anisotropic bond-dependent exchange couplings as
\begin{eqnarray}
J_{ij}^{xx} & = & -2\left(J_{\pm}-J_{\pm\pm}\mbox{cos}\theta_{ij}\right),\label{eq:jxx}\\
J_{ij}^{yy} & = & -2\left(J_{\pm}+J_{\pm\pm}\mbox{cos}\theta_{ij}\right),\label{eq:jyy}\\
J_{ij}^{xy} & = & -2J_{\pm\pm}\mbox{sin}\theta_{ij}.\label{eq:jxy}
\end{eqnarray}

For $-2\le J_{\pm\pm}/J_{\pm}<0$ thermal (and quantum) fluctuations select the state $\psi_{3}$, whereas in the region $0<J_{\pm\pm}/J_{\pm}\le2$ the state $\psi_{2}$ is selected. We are primarily interested in the situation where fluctuations and quenched disorder tend to select {\em different} states,\citep{maryasin14,andreanov15} hence we are going to consider $J_{\pm\pm}=J_{\pm}$ in our simulations described below.

\subsection{Monte-Carlo algorithm}

We perform classical Monte-Carlo simulations of the model \ref{eq:spin_pyro_xy_aniso} for clusters with $N=4L^{3}$ spins with periodic boundary conditions. To equilibrate the system efficiently we perform three types of Monte-Carlo moves. In addition to single-site Metropolis moves, we consider microcanonical steps to improve the sampling at lower temperatures.\citep{alonso96} We consider the ratio of $10$ microcanonical sweeps to each Metropolis sweep.\citep{pixley08} One Monte-Carlo sweep corresponds to one attempted spin flip at each of the $N$ lattice sites. Typically, we perform $\sim10^{5}$ Monte-Carlo sweeps as initial thermalization followed by further $\sim10^{5}$ sweeps to obtain thermal averages, which are calculated by dividing the measurement steps into $10$
bins.
We also do ``parallel tempering'' sweeps,\citep{partemp_jpsj} which are necessary to prevent the system being trapped in a valley in configuration space at low temperatures. We do one parallel tempering sweep after each Metropolis sweep. We note that parallel tempering limits the maximum system size to $L=12$. In all our results we set the Boltzmann constant $k_{B}=1$ and use $J_{\pm}$ as our energy scale.

To monitor the order in the local XY planes we compute
\begin{equation}
m_{x}=\frac{1}{N}\sum_{i=1}^{N}S_{i}^{x}, ~~ m_{y}=\frac{1}{N}\sum_{i=1}^{N}S_{i}^{y},\label{eq:mx_my}
\end{equation}
where the spins components are calculated in the \emph{local} frame. The order parameter to detect in-plane magnetic order can then be defined as
\begin{equation}
m=\sqrt{m_{x}^{2}+m_{y}^{2}},\label{eq:m_def}
\end{equation}
to which we also associate a magnetic correlation length $\xi^{\perp}$, calculated from the corresponding static structure factor. The order parameter $m$ cannot distinguish between the states $\psi_{2}$ and $\psi_{3}$. To discriminate between them we define a clock-like order parameter
\begin{equation}
m_{6}=m\cos (6\theta),\label{eq:m6_def}
\end{equation}
where $\theta=\tan^{-1}\left(m_{y}/m_{x}\right)$. We then have $m_{6}>0$ for the six noncoplanar $\psi_{2}$ states and $m_{6}<0$ for the coplanar $\psi_{3}$ states.

On general grounds, the ordering in the model \ref{eq:spin_pyro_xy_aniso} can be expected to correspond to that of a 3D XY model with a $Z_{6}$ anisotropy term. At criticality, this anisotropy is dangerously irrelevant, and critical exponents are identical to that of an isotropic XY model. However, the anisotropy is relevant for $T<T_{c}$ and lengths larger than $\Lambda\propto{(\xi^{\perp})}^{\alpha_6}$ where $\alpha_6>1$ is an exponent characterizing the scaling dimension of the anisotropy.\citep{zhitomirsky14,lou07,wenzel11}
Therefore, \emph{in a finite-system simulation}, the order-by-disorder selection mechanism only takes place for $L\gtrsim\Lambda$. Therefore we either have to go to low temperatures or large systems. We thus expect\citep{zhitomirsky14} strong finite-size effects in $m_{6}$. In order to deal with those we also calculate histograms $P\left(m_{6}\right)$ of $m_{6}$ which we expect to be strongly peaked around an either positive or negative average at sufficiently low $T$.

\subsection{Clean system}
\label{sec:clean}

\begin{figure}[t]
\begin{centering}
\includegraphics[scale=0.3]{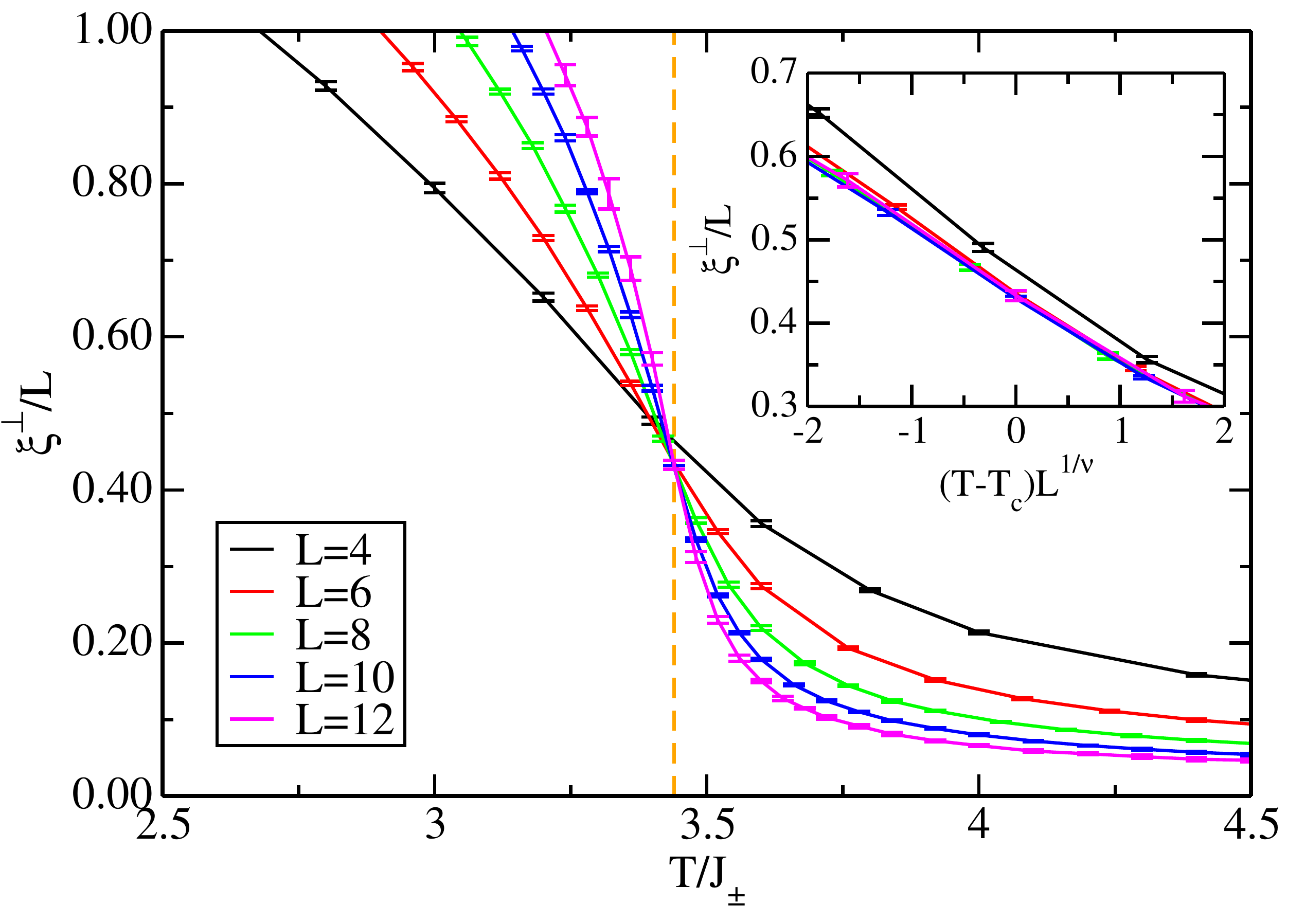}
\end{centering}

\caption{\label{fig:xi_clean}Results for the clean XY pyrochlore model
with $J_{\pm\pm}=J_{\pm}$. In-plane
magnetic correlation length $\xi^{\perp}$, divided by the system
size $L$, as a function of the temperature $T$. The curves for different
system sizes all cross at the critical point. Inset: Scaling of the
different curves using the 3D XY correlation length exponent. The vertical dashed line marks the position of the critical temperature $\Tc/J_{\pm}=3.44\left(1\right)$. }
\end{figure}

To test our simulation code, we have first studied the clean system and compared the results with those available in the literature. To detect the phase transition we study the crossing points of the in-plane magnetic correlation length, $\xi^{\perp}/L$, Fig.~\ref{fig:xi_clean}. We obtain $\Tc/J_{\pm}=3.44\left(1\right)$ which also yields a very good data collapse
--- apart from the smallest system size $L=4$ --- with the correlation-length exponent $\nu=0.672$, the best estimate for
the 3D XY universality class.\citep{campostrini06}
This value of $\Tc$ also compares well with that of Ref.~\onlinecite{zhitomirsky14}: In their units, our $\Tc$ is given by $\Tc=0.65$, to be compared to their value of $0.665$.
We note that the Curie-Weiss temperature for this choice of parameters is $\Tcw=4J_{\pm}$. We thus see that, although the systems is very frustrated, the ratio $\Tc/\Tcw=0.86\sim O\left(1\right)$.

\begin{figure}[!t]
\includegraphics[scale=0.3]{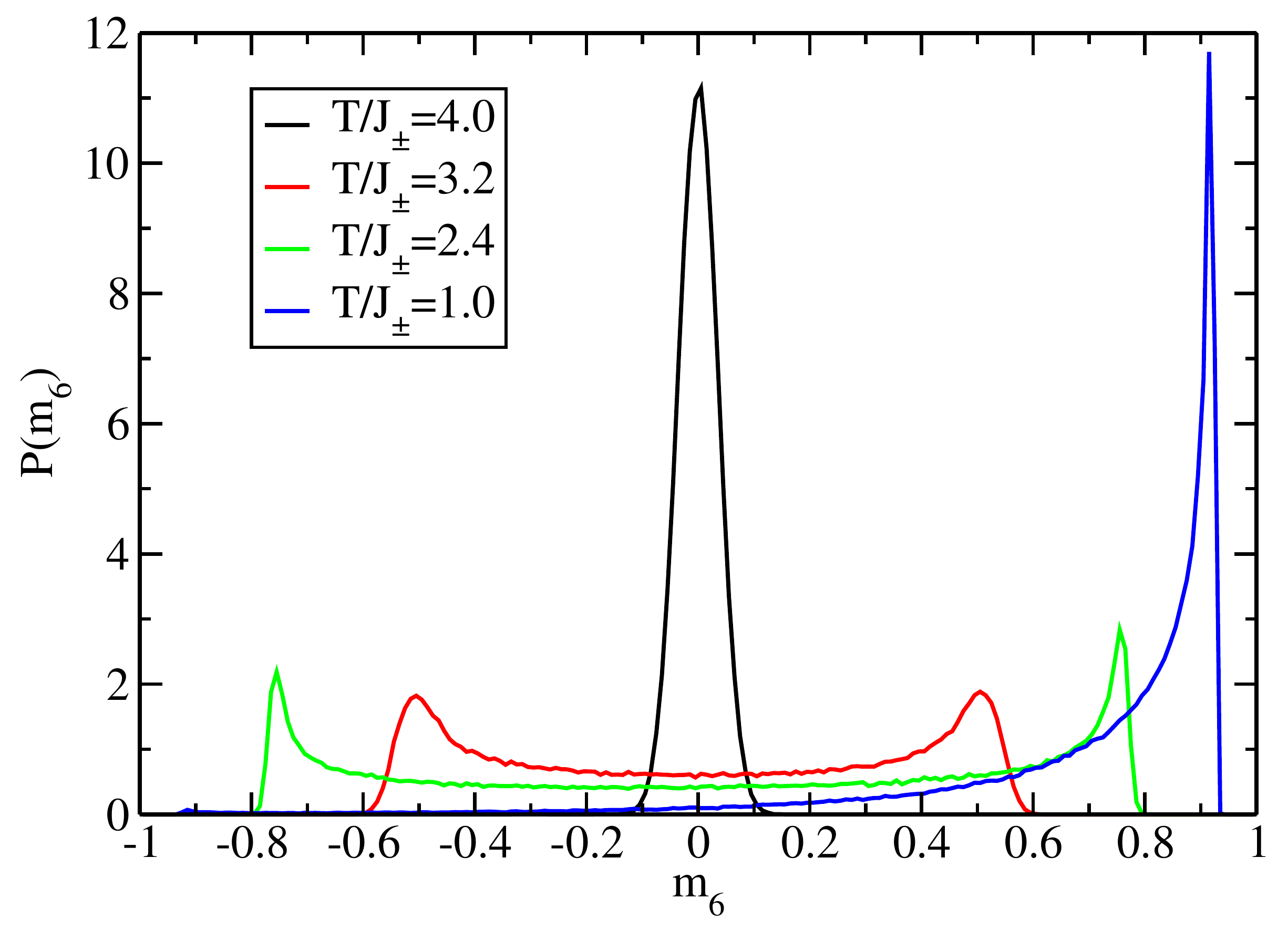}
\caption{\label{fig:hist_m6_clean}$m_{6}$ histograms for the clean XY
pyrochlore model with $J_{\pm\pm}=J_{\pm}$ for $L=10$, at different values of temperatures
$T$. The critical temperature is $\Tc/J_{\pm}=3.44\left(1\right)$.}
\end{figure}

We have also studied the order parameters $m$ and $m_6$ (not shown), with results which are again in agreement with those of Ref.~\onlinecite{zhitomirsky14}. Histograms of $m_{6}$, illustrating the finite-size crossovers, are shown in Fig. \ref{fig:hist_m6_clean}. For $T>\Tc$, $P\left(m_{6}\right)$ is peaked around $m_{6}=0$, with the width of the peak proportional to $1/N$. For $T\lesssim \Tc$ the distribution becomes bimodal indicating a superposition of $\psi_{2}$ and $\psi_{3}$. Upon lowering $T$ it finally peaks around a positive value of $m_{6}$ once $L\gtrsim\Lambda$ -- this is the signature of the $\psi_2$ state being selected from the energetically degenerate manifold of states.

\subsection{Disordered system}

We now introduce bond disorder into the system, with the main motivation to understand the spin freezing in {\nccf}. We choose
\begin{eqnarray}
J_{\pm}^{ij} & = & J_{\pm}\left(1+\epsilon_{ij}\right),\label{eq:disorder_jp}\\
J_{\pm\pm}^{ij} & = & J_{\pm\pm}\left(1+\epsilon_{ij}\right),\label{eq:disorder_jpp}
\end{eqnarray}
where $\epsilon_{ij}$ is a random variable uniformly distributed between $\pm W/2$. Notice that we use the same $\epsilon_{ij}$ for both $J_{\pm}^{ij}$ and $J_{\pm\pm}^{ij}$. Other forms of disorder (e.g. bimodal) are not expected to change the results qualitatively. Given our limited system sizes, we choose to consider strong disorder, $W=2$, the maximum value of disorder for which the sign
of $J_{\pm\pm}^{ij}$ does not change. The Monte-Carlo simulations now also involve averaging over $N_{rlz}$ realizations of disorder; here we have employed values ranging from $N_{rlz}=400$ for $L=5$ to $N_{rlz}=21$ for $L=10$.

\begin{figure}[t]
\begin{centering}
\includegraphics[scale=0.3]{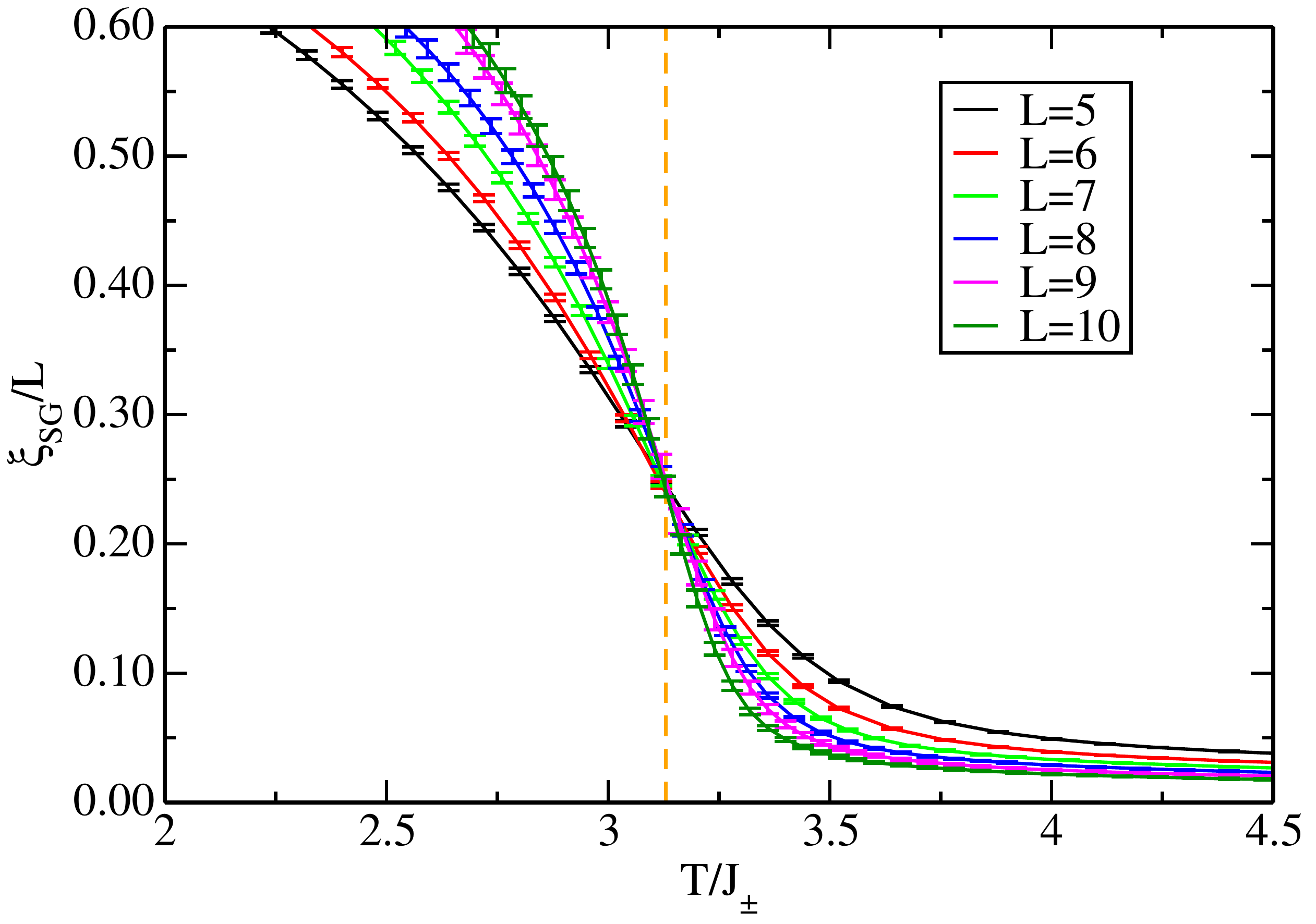}
\includegraphics[scale=0.3]{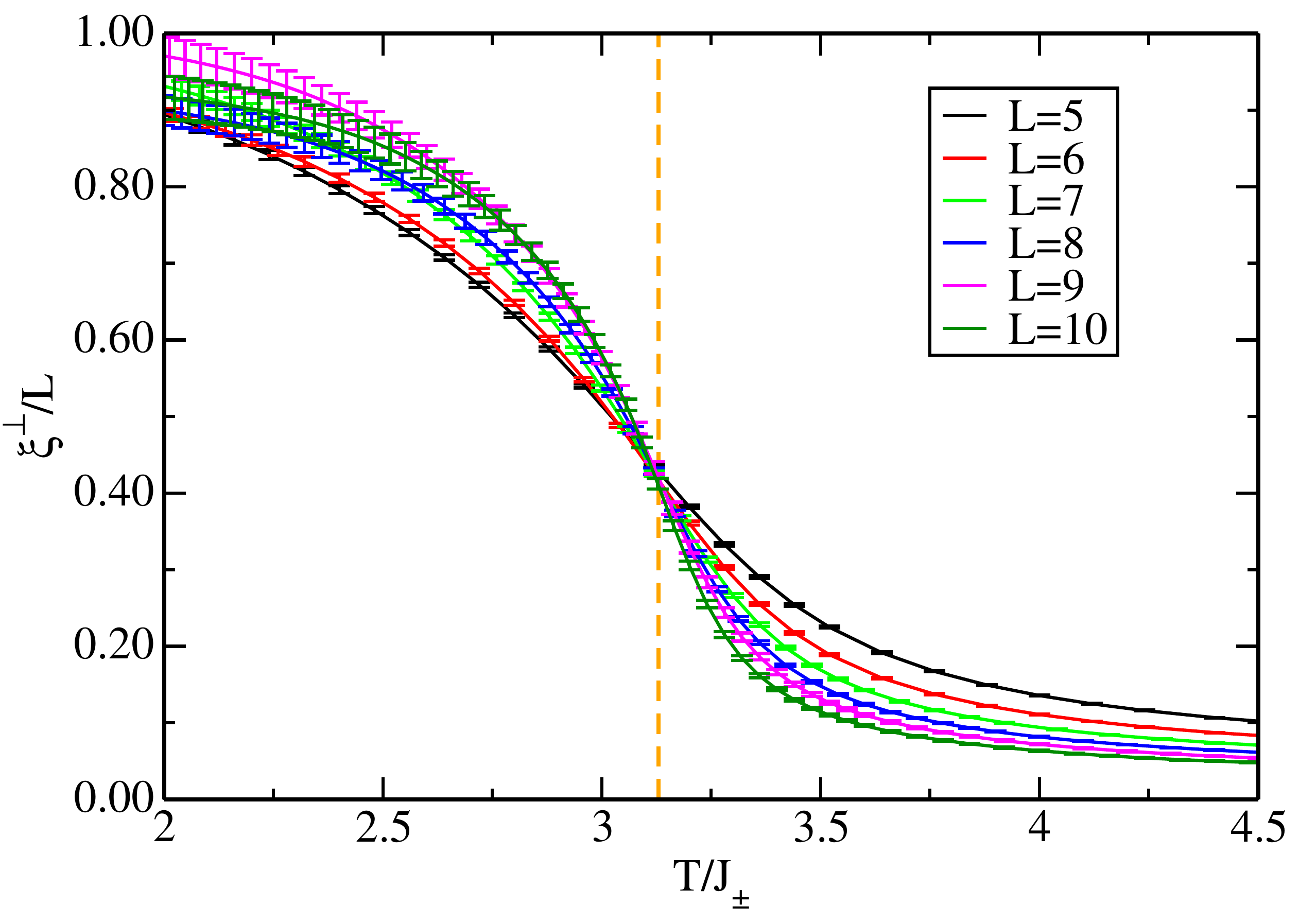}
\end{centering}

\caption{\label{fig:xisg_xil_w20}Results for the disordered XY pyrochlore
model with $J_{\pm\pm}=J_{\pm}$ and $W=2$. Top: Spin-glass correlation
length. Bottom: In-plane magnetic correlation length $\xi^{\perp}$,
divided by the system size $L$, as a function of the temperature
$T$. The vertical dashed line marks the position of the freezing
temperature $\Tf/J_{\pm}=3.15\left(2\right)$ (top) and $\Tf/J_{\pm}=3.13\left(2\right)$
(bottom). }
\end{figure}



Anticipating spin-glass physics, we have calculated both the magnetic correlation length $\xi^{\perp}$ and the spin-glass correlation length $\xi_{SG}$, defined as the correlation length of the Edwards-Anderson order parameter. The latter captures any kind of spin freezing, even if the system
shows no long-range magnetic order. As we show in Fig. \ref{fig:xisg_xil_w20}, we can extract a freezing temperature $\Tf/J_{\pm}=3.15\left(2\right)$ by the crossing point of $\xi_{SG}/L$. The scaled magnetic correlation length $\xi^{\perp}/L$ also shows a (approximate) crossing point at $\Tf/J_{\pm}=3.13\left(2\right)$. Compared to the clean-system result, Fig.~\ref{fig:xi_clean}, it is apparent that the correlation length $\xi^{\perp}$ is reduced by disorder, but it is still comparable to our largest system size. However, the systematic downward shift of $\xi^{\perp}/L$ at low $T$ hints at the absence of long-range order in the disordered system.

The influence of bond disorder is best seen in the histograms of $m_6$, Fig. \ref{figmc} (lower panel). Unlike in the clean case, here $P\left(m_{6}\right)$ remains bimodal down to the lowest simulation temperatures. Correspondingly, the averaged $m_6$ values are close to zero (not shown). These results show that the system contains coexisting fluctuating domains of $\psi_{2}$ and $\psi_{3}$ states. The occurrence of local $\psi_3$ configurations is consistent with the perturbative arguments put forward in Ref.~\onlinecite{maryasin14}. However, we see indications neither for long-range order of $\psi_{3}$ type being selected by bond disorder nor for a thermal first-order transition between $\psi_{2}$ and $\psi_{3}$ phases.\citep{maryasin14}

Considering the apparent discrepancy between our results and the conclusions of Refs.~\onlinecite{maryasin14,andreanov15} we note the following:
(a) Previous finite-temperature Monte-Carlo results in the presence of quenched disorder have only been reported for spin dilution,\citep{maryasin14} whereas we have studied the case of (strong) bond disorder. Although analytical arguments \citep{maryasin14} suggest the two cases to be qualitatively similar, this is not guaranteed.
(b) The numerical simulations of Refs.~\onlinecite{maryasin14,andreanov15} do not appear to take into account the possibility of glassiness. For the latter, parallel tempering is inevitable to avoid trapping in local minima in configuration space.
Given that glassiness is a rather natural scenario for spin systems with frustration and disorder, our results suggest to revisit the XY model with spin dilution. Work in this direction is in progress.

\subsection{Discussion}

Our Monte-Carlo simulations show that strong bond disorder in the pyrochlore XY antiferromagnet can produce a low-temperature state which is best described as a glassy mixture of $\psi_{2}$ and $\psi_{3}$ configurations. While such a state appears qualitatively consistent with the experimental data on {\nccf}, there are significant quantitative discrepancies:
(i) Even with strong bond disorder, $W=2$, the magnetic correlation length in the model calculation remains sizeable, $\xi^{\perp} \gtrsim 10$. In contrast, neutron scattering indicates \cite{Kate-Ross2015} $\xi^{\perp} \lesssim 2$.
(ii) The model calculations yield $\Tcw/\Tf \approx 1.3$, to be contrasted with the experimental value of $56$.

Together, this shows that the model -- even with strong quenched disorder implemented -- overestimates the tendency to spin order, i.e., underestimates fluctuation effects. Two obvious sources come to mind, namely quantum fluctuations and fluctuations out of the XY manifold of states due to small crystal-field splitting. For the latter, neutron-scattering data suggest \cite{Kate-Ross2015} a relevant crystal-field scale to be $2.5$\,meV -- this is not huge, but still significantly larger than $\Tf$ and indicates that the XY approximation may be justified. Still, the interplay of quenched disorder and a small crystal-field splitting may lead to non-trivial physics which is unexplored.

The effect of quantum fluctuation is difficult to estimate. In this context, it is interesting to note that {\ertio} displays a frustration index $\Tcw/\Tc \approx 18$ (Ref.~\onlinecite{champion03}) while the classical XY model yields $1.2$ (Section \ref{sec:clean}). Notably, a series-expansion study \cite{oitmaa13} of a full quantum Hamiltonian, with parameters extracted from high-field neutron scattering results,\cite{savary12} reproduces the correct value of $\Tc\approx 1.2$\,K, hence quantum fluctuations account for a drastic reduction of the ordering temperature. Applied to {\nccf} which has an even larger frustration index, we consider it possible that quantum fluctuations bring the system close to a quantum spin-liquid regime. The effect of quenched disorder in such a spin-liquid regime is a fascinating topic for future studies.

\bibliography{NaCaCoF_PRB}

\end{document}